\pdfoutput=1

\documentclass[journal,onecolumn,11pt]{IEEEtran}
\usepackage{setspace}
\usepackage{hyperref}
\hypersetup{
    colorlinks=true,
    linkcolor=blue,
    filecolor=magenta,      
    urlcolor=cyan,
}

\usepackage{siunitx}
\usepackage{cite}

\ifCLASSINFOpdf
   \usepackage[pdftex]{graphicx}
\else
\fi
\usepackage{amsmath}

\newcommand{\vect}[1]{\mathbf{#1}}
\usepackage{url}

\begin{document}
%
\title{Real-time interactive 4D-STEM phase-contrast imaging from electron event representation data}

\author{Philipp M. Pelz, Ian Johnson, Colin Ophus, Peter Ercius, Mary C. Scott}

\maketitle
\begin{abstract}
The arrival of direct electron detectors (DED) with high frame-rates in the field of scanning transmission electron microscopy has enabled many experimental techniques that require collection of a full diffraction pattern at each scan position, a field which is subsumed under the name four dimensional-scanning transmission electron microscopy (4D-STEM). DED frame rates approaching 100 kHz require data transmission rates and data storage capabilities that exceed commonly available computing infrastructure. Current commercial DEDs allow the user to make compromises in pixel bit depth, detector binning or windowing to reduce the per-frame file size and allow higher frame rates. This change in detector specifications requires decisions to be made before data acquisition that may reduce or lose information that could have been advantageous during data analysis. The 4D Camera, a DED with 87 kHz frame-rate developed at Lawrence Berkeley National Laboratory, reduces the raw data to a linear-index encoded electron event representation (EER). Here we show with experimental data from the 4D Camera that linear-index encoded EER and its direct use in 4D-STEM phase contrast imaging methods enables real-time, interactive phase-contrast from large-area 4D-STEM datasets. We detail the computational complexity advantages of the EER and the necessary computational steps to achieve real-time interactive ptychography and center-of-mass differential phase contrast using commonly available hardware accelerators.
\end{abstract}

\begin{IEEEkeywords}
transmission electron microscopy, 4D-STEM, phase contrast imaging, sparse data representation
\end{IEEEkeywords}

\IEEEpeerreviewmaketitle

\section{Introduction}

Technological developments in electron microscopy continuously advance the frontier of observable phenomena. Ultimately, we would like this frontier to be defined by fundamental laws and not our instrumentation's technological shortcomings within a reasonable budget. \\
In the last decade, the technical development of DEDs has enabled a number of scientific breakthroughs in electron microscopy. In cryo-electron microscopy (cryo-EM), DED technology and advanced algorithms have enabled atomic-resolution imaging of biological macromolecules \cite{nakane_single-particle_2020,Yip_Fischer_Paknia_Chari_Stark_2020}. DED technology has helped to surpass resolution limits set by electron optics in materials science \cite{Jiang_2018, Chen_2020} and image light and heavy elements simultaneously at atomic resolution and low electron dose \cite{Yang_2016,Lozano_2018,OLeary_2020} in a family of techniques known as four dimensional-scanning transmission electron microscopy (4D-STEM) \cite{Ophus_2019}. Yet, as internal DED frame-rates approaching \SI{2.5}{\kilo\hertz} for large cryo-EM detectors and \SI{100}{\kilo\hertz} for smaller 4D-STEM detectors enable readout rates on the order of $10^{10}$ pixels per second, the previous strategy of transmitting and storing uncompressed recorded data is being overhauled to enable large-scale, extended experiments.
\begin{figure*}[h]
    \centering
        \includegraphics[width=\textwidth]{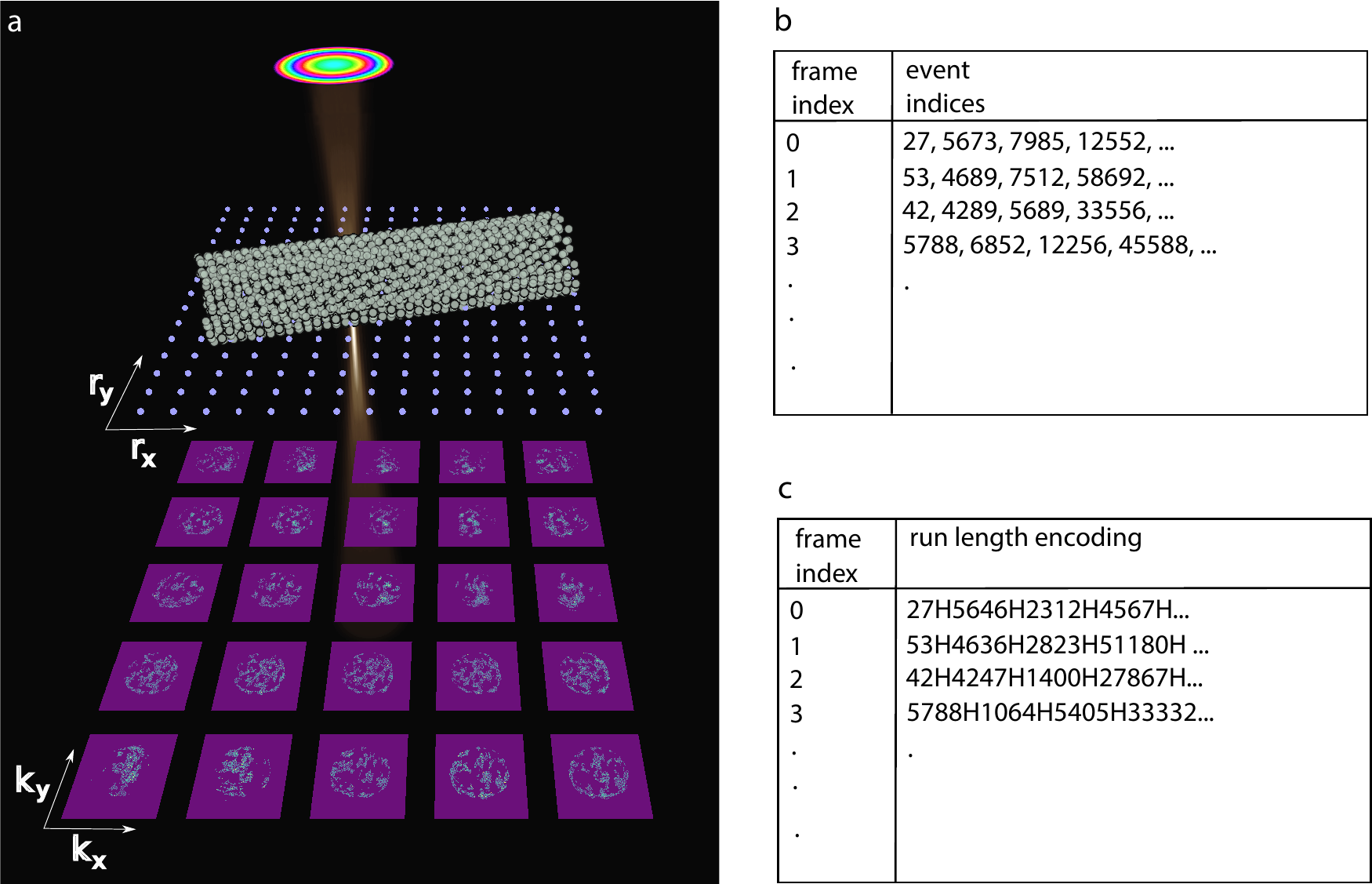}
    \caption{a) Sketch of a 4D-STEM experiment with sparse electron counts. A coherent electron beam is scanned over the sample, here a carbon nanotube, in a two-dimensional lattice in real-space, marked by the blue grid of points. In diffraction space, the detector records a two-dimensional image for each scanned point. For high frame rates close to \SI{100}{\kilo\hertz} and low-dose conditions, the number of electrons per recorded frame is much smaller than the number of pixels, and the data can be efficiently compressed. b) Electron-event representation in linear-index encoding. Each frame is converted to a list of electron hit coordinates. The memory requirements grow linearly with the electron counts. c) Electron-event representation with run-length encoding, where only the linear distance between the electron hits is saved.}
	\label{fig:sketch}
\end{figure*}

\section{Electron event format for 4D-STEM data and implications for signal processing}

Until recently, the connected hardware's data-transfer capabilities limited the frame-rate of DEDs and often the size of the recorded datasets \cite{Guo_2020,Datta_2020}. The reason for this limitation is that the sensor data was transferred uncompressed over network to data-processing hardware, such that the possible frame-rate was intrinsically limited by the data transfer rate. A straightforward way to remove this data-transfer bottleneck is to reduce and/or compress the data as they arrive in memory, and strategies to this end have been developed recently for DEDs \cite{Guo_2020,Datta_2020}.
DEDs such as the detectors employed in cryo-EM and the 4D Camera are operated with experimental parameters that naturally allow data reduction and compression by only recording a limited number of electrons per frame. If the number of electrons per pixel per frame is limited to 1/100 to 1/40 of the total detector pixels to avoid coincidence losses \cite{Guo_2020}, the position of electron hits can be determined by extraction of clusters of deposited energy in raw frames and subsequent centroiding to determine the most likely to initial electron strike position \cite{battaglia_cluster_2009}.
Once electron event coordinates are determined, they can be further encoded to reduce storage and requirements. Three different encoding strategies have been proposed: First, direct encoding as a linear index into the camera coordinates, second, run-length encoding of the pixel distances between electron hits, and third, binary encoding as a 2D binary array \cite{Guo_2020,Datta_2020}. The 4D Camera software currently implements linear index encoding, while the Falcon3 and 4 cameras implement run-length encoding.
\section{Generalized diffraction pattern analysis algorithms}
In this section we briefly discuss common 4D-STEM processing steps such as binning and cropping, which are accelerated by the linear index representation. 

In 4D-STEM data streams, binning can be performed in real space, in diffraction space, or both. Binning is commonly used to reduce the diffraction space sampling for applications where finer diffraction-space sampling increases the memory requirements without improving measurement precision or resolution. This is usually the case when the sample is relatively thin and position in the focus of the electron beam, such that features in the diffraction pattern are large and can be sampled with only few pixels. Such applications include scanning nanodiffraction, direct ptychography solvers like single-sideband reconstruction and Wigner-Distribution deconvolution \cite{Yang_Pennycook_Nellist_2015}, iterative solvers such as those used in electron ptychography algorithms \cite{Humphry_Kraus_Hurst_2012,Chen_2020} and S-matrix inversion from full diffraction patterns \cite{Pelz_Brown_Ciston_Findlay_Zhang_Scott_Ophus_2020}.\\
Cropping is usually used in techniques operating on whole diffraction patterns, like ptychography or S-matrix inversion, to center the diffraction pattern or discard data that is not modeled by the chosen reconstruction method.\\
Center of mass measurements are used in differential phase contrast imaging. Radial sums are used to calculate virtual annular detectors, and sums of multiple frames are used to create position-averaged convergent beam electron diffraction (PACBED) images, and to extract average vacuum probes. The computational complexity of these operations is shown in Table \ref{tab:complexity}, where $\mathsf{M}$ is the number of detector pixels, $\mathsf{C}$ is the number of electron counts, $\mathsf{F}$ is the number of frames, and $\mathsf{r}_1$ and $\mathsf{r}_2$ are the inner and outer radii of integration.
\begin{table}[h!]
\begin{center}
\begin{tabular}{|c|c|c|c|} 
\hline
  & dense format & linear index encoding & run-length encoding\\ 
 \hline\hline
 bin & ${O(\mathsf{M}\cdot\mathsf{F})}$ & ${O(\mathsf{C}\cdot\mathsf{F})}$& ${O(2\mathsf{C}\cdot\mathsf{F})}$ \\ 
 \hline
  crop & ${O(\mathsf{M}\cdot\mathsf{F})}$ & ${O(\mathsf{C}\cdot\mathsf{F})}$& ${O(2\mathsf{C}\cdot\mathsf{F})}$ \\ 
 \hline
 center of mass & ${O(\mathsf{M}\cdot\mathsf{F})}$ & ${O(\mathsf{C}\cdot\mathsf{F})}$& ${O(2\mathsf{C}\cdot\mathsf{F})}$ \\ 
 \hline
 radial sum & ${O(\mathsf{r}_1^2-\mathsf{r}_2^2)\cdot\mathsf{F}}$ & ${O(\mathsf{C}\cdot\mathsf{F})}$& ${O(2\mathsf{C}\cdot\mathsf{F})}$ \\ 
  \hline
 sum all frames & ${O(\mathsf{M}\cdot\mathsf{F})}$ & ${O(\mathsf{C}\cdot\mathsf{F})}$& ${O(2\mathsf{C}\cdot\mathsf{F})}$ \\ 
 \hline
\end{tabular}
\end{center}
\caption{Computational complexity of common operations in 4D-STEM with different encoding schemes, for $\mathsf{M}$ detector pixels, $\mathsf{F}$ image frames, $\mathsf{C}$ electron counts, and an $\mathsf{r}_1$ to $\mathsf{r}_2$ radial range.}
\label{tab:complexity} 
\end{table}
Many operations scale with the number of counts when linear index encoding or run-length encoding are used, whereas they scale with the number of pixels when a dense data format is used. Re-coding run-length encoded data into linear-index encoded format requires ${O(\mathsf{C})}$ operations, therefore the run-length encoded format is only slightly more inefficient for common pre-processing operations in 4D-STEM, while offering a compression advantage \cite{Datta_2020}. Given that the number of counts is 40 to 100 times lower than the number of pixels in back-thinned detectors \cite{Guo_2020}, this results in a 40 to 100-fold speedup in computation time for these operations. 
In the following sections we describe how we make use of linear index encoding and an efficient implementation on commodity hardware accelerators to achieve real-time electron ptychography and DPC reconstructions from 4D-STEM data in linear index encoded EER format.
\section{Real-time, interactive electron ptychography}
Ptychography is a general method to retrieve the structure of thin samples from scanning diffraction measurements under the projection approximation, and to achieve super-resolution beyond the capabilities of focusing optics \cite{Rodenburg_Bates_1992,Nellist_McCallum_Rodenburg_1995,Nellist_Rodenburg_1994,Jiang_2018}. As a data-intensive computational imaging technique that requires scanning diffraction measurements, it has only recently achieved its full potential in electron microscopy with the arrival of DEDs. Recent advances include record resolution \cite{Jiang_2018} and low-dose phase-contrast imaging of 2D materials \cite{Chen_2020}, perovskites \cite{Reis_Yang_Ophus_Ercius_Bizarri_Perrodin_Shalapska_Bourret_Ciston_Dahmen_2018},  biological macromolecules \cite{Pelz_Bucker_Ramm_Venugopal_Kassier_Eggert_Lu_DuninBorkowski_Miller_2019} and viruses \cite{Zhou_Song_Kim_Pei_Huang_Boyce_Mendona_Clare_Siebert_Allen_2020}. The high frame-rate of the 4D Camera allow acquisition of large-area scans with 1024 x 1024 scan positions in under 15 seconds, and the small file sizes of compressed 4D-STEM data in EER format (currently ca. 8 minutes post-acquisition processing time) allow reconstruction with the direct single-sideband reconstruction in under 1 second. The single-sideband reconstruction \cite{rodenburg1993experimental} is attractive for providing direct feedback during the 4D-STEM experiment session because of its low computational complexity. We detail the reconstruction algorithm below.
\begin{figure*}[h]
    \centering
        \includegraphics[width=\textwidth]{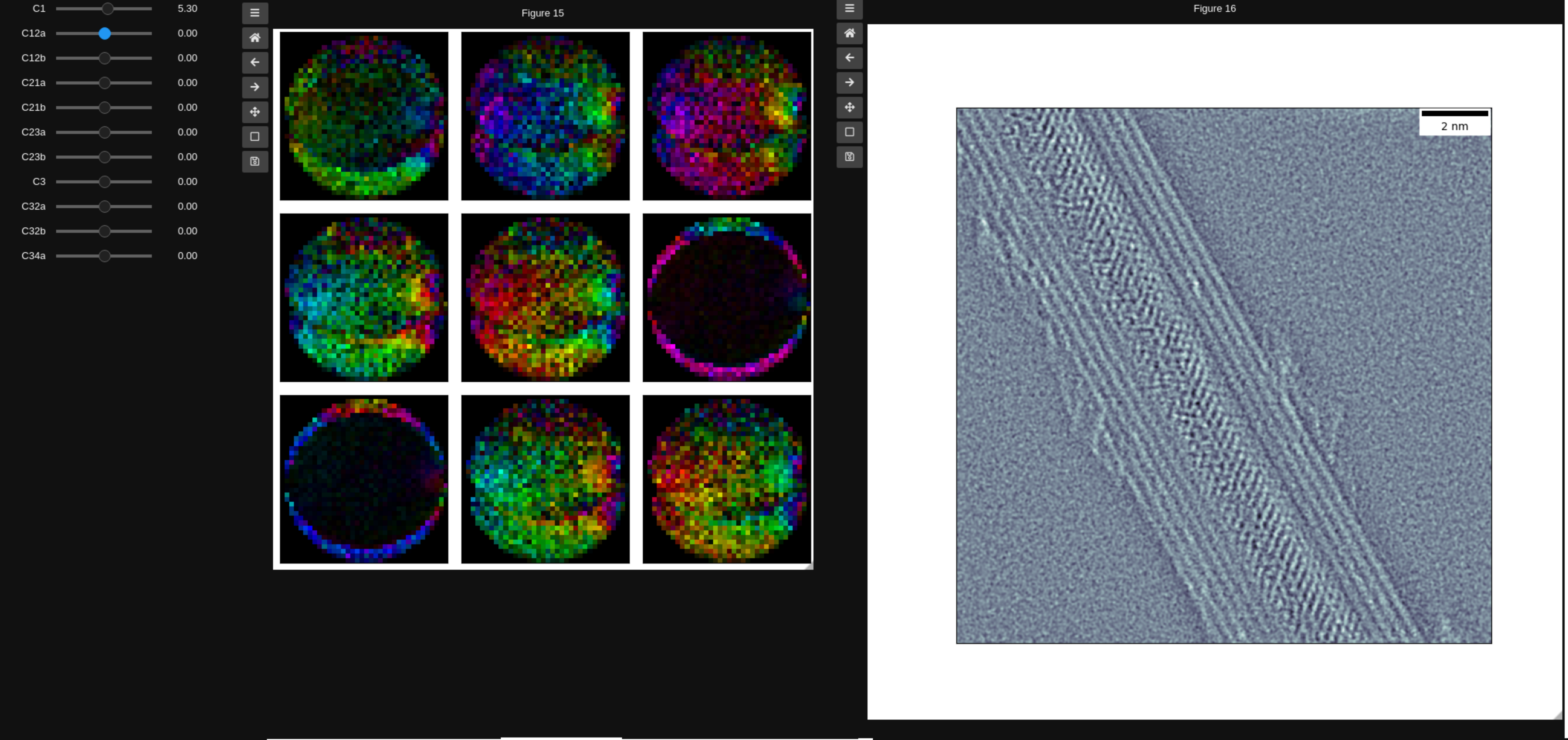}
    \caption{Screenshot of interactive ptychography reconstruction from 4D-STEM data in EER format. The panel on the left shows tunable aberration coefficients. The panel in the middle shows nine selected spatial frequencies of the $G$ function for assessment of aberrations. The panel on the right updates with sub-second response time the ptychography reconstruction given the aberrations chosen on the left. This enables to perform virtual focusing through the sample in real time.}
	\label{fig:ssb}
\end{figure*}
Beginning from the projection approximation, the measured intensity $I$ at a spatial frequency $\mathbf{q}$ and a probe position $\mathbf{R}$ is given by
\begin{equation}
    I(\mathbf{k},\mathbf{R}) = \left|\int a(\mathbf{r}-\mathbf{R})T(\mathbf{r}) e^{2 \pi i \mathbf{r}\cdot\mathbf{k}} d\mathbf{r}\right|^2,
\end{equation}
where $a(\mathbf{r})$ is the complex probe function in real space, and $T(\mathbf{r})$ is the complex transmission function. The goal is to retrieve $T$ from $I$.
We first consider the transformed datacube $G(\mathbf{k},\mathbf{K}) = \mathcal{F}_{\mathbf{R}\to\mathbf{K}}I(\mathbf{k},\mathbf{R})$, where $\mathbf{K}$ is the reciprocal coordinate to scan position $\mathbf{R}$. Assuming the sample is a weak phase object, this may be written as \cite{rodenburg1993experimental}
\begin{equation}
\label{equ:g_function}
    G(\vect{k},\vect{K}) = \left|A(\vect{k})\right|^2\delta(\mathbf{k}) 
    + A(\vect{k})A^*(\vect{k}+\vect{K})T(-\mathbf{k})^* 
    -A^*(\mathbf{k})A(\mathbf{k}-\mathbf{K})T(\mathbf{k})
\end{equation}
where $T(\mathbf{k}) = \mathcal{F}_{\mathbf{r}\to\mathbf{k}}T(\mathbf{r})$.
The latter two terms in Eq.~\ref{equ:g_function} each contain the aperture function $A(\vect{k})$, one centered at $0$ and one shifted by the scan spatial frequency $\mathbf{K}$.
Since $A(\vect{k})$ is bandwidth-limited by the condenser aperture, we can eliminate one of the two last terms in Eq.~\ref{equ:g_function} by only considering points lying in the overlap region of two of the three shifted apertures $A(\vect{k})$, $A(\vect{k}+\vect{K})$ and $A(\vect{k}-\vect{K})$.
To eliminate the second term, we define the set of pixels
\begin{equation*}
    \mathcal{K} = \{\mathbf{q}\,:\,(\lvert\mathbf{q}\rvert < k_0) \land\,(\lvert\mathbf{q}+\mathbf{q}\rvert > k_0) 
    \land\,(\lvert\mathbf{q}-\mathbf{q}\rvert < k_0)\}o
\end{equation*}
where $\land$ is the logical and operation and the maximum disk size is $k_0 = \frac{\alpha}{\lambda}$ for convergence semi-angle $\alpha$ and electron wavelength $\lambda$.
If only data from $\mathbf{q} \in \mathcal{K}$ is used, the second term in Eq.~\ref{equ:g_function} vanishes, so that 
\begin{equation}
\label{equ:t_recon}
    T(\mathbf{q}) = \sum_{\mathbf{q} \in \mathcal{K}} \frac{G(\mathbf{q},\mathbf{q})}{A(\vect{q})A^*(\mathbf{q}-\mathbf{q})}.
\end{equation}
Since the method uses only data from one of the double-overlap regions, the method has been named single-sideband (SSB) ptychography. This method can be implemented very efficiently, because it only requires a batched 2D Fast Fourier Transform (FFT) along the real-space coordinates of the dataset and a complex multiplication and reduction to 2D. These operations can both be implemented efficiently with a complex matrix multiplication, or a custom CUDA kernel. For changing the virtual focus of the reconstruction, only the multiplication-reduction step in Eq.~\ref{equ:t_recon} has to be performed again, and only a small additional amount of memory for the 2D aperture function has to be stored.
For ease of implementation, we have chosen to implement a custom cuda kernel for the operation in Eq.~\ref{equ:t_recon}, although a complex matrix multiplication could offer additional performance benefits with little memory overhead. 
We implement SSB ptychography from linear index EER in the following way: using the insight that sampling the bright-field disk of a moderately aberrated electron probe with more than 32x32 pixels leads to no improvements in reconstruction quality \cite{Yang_Pennycook_Nellist_2015}, we load the EER data directly on the GPU and perform centering, cropping and binning to a diffraction pattern size smaller than \num{32 x 32} pixels directly from the EER format. We then densify the binned and cropped data to prepare for the Fourier transform along the scan coordinate axis.
The resulting dense 4D-STEM dataset usually fits on currently available GPUs with \num{24} GB or more working memory, such that no inter-device transfers are necessary. \\
We then perform a batched 2D Fourier transform along the scan coordinate to compute $G(\vect{k},\vect{K})$. The subsequent summation over the reciprocal scan coordinates is implemented in a custom cuda kernel with the numba package, with the mask $\mathcal{K}$ and the aperture functions computed inline, such that no additional memory is required. The total computation time without memory transfer to GPU for a dataset with \num{1024 x 1024} probe positions cropped to \num{18 x 18} pixels in the detector coordinates is roughly 50 ms. This is fast enough that aberration corrections can be applied interactively, for example to tune the virtual focus of the dataset. The algorithm is implemented in the Python language using the cupy package for general array manipulation and the numba package for just-in-time compilation of custom cuda kernels. A graphical frontend is implemented using the ipywidgets and matplotlib packages in jupyter notebook, such that the reconstruction can be run remotely on a compute cluster, with a local graphical user interface in the browser. 
\section{Real-time differential phase contrast}
While real-time differential phase-contrast (DPC) using specialized annular detectors has been already in use for a while \cite{Shibata_2012,Yucelen_2018}, center of mass DPC from pixelated detectors can offer a sensitivity advantage \cite{Muller_2014,Yang_Pennycook_Nellist_2015}. Compared to electron ptychography, DPC has low memory requirements, since the reduction step of computing the center of mass field in x- and y-directions can be performed directly from the data in linear-index EER format, and the probe is usually assumed to be aberration-free. Following \cite{close2015towards,lazic2016phase}, we compute the center-of-mass images from the EER dataset,
\begin{equation}
\label{E:dpc_detector}
    \mathbf{I}(\mathbf{R}) = \int \lvert \Psi(\mathbf{q},\mathbf{R}) \rvert^2 \mathbf{q} d\mathbf{q},
\end{equation}
which is a simple integration over momentum space and can be performed in ${O(\mathsf{C})}$ computation time with the data in linear-index EER format. The electrostatic potential can then be obtained iteratively by solving the least-squares problem
\begin{equation*}
    \operatorname*{arg\,min}_{\phi}\displaystyle\mathcal{L}(\phi) = \left|\left|\partial_{\mathbf{R}} \phi - \mathrm{COM}\right|\right|_2,
\end{equation*}
where $\phi$ is the reconstructed DPC image, and $\mathrm{COM}$ is the measured center-of-mass image. We solve the problem with a simple regularized gradient descent routine as \cite{Savitzky_Hughes_Zeltmann_Brown_Zhao_Pelz_Barnard_Donohue_DaCosta_Pekin_2020}:
\begin{equation*}
    \phi^{\mathsf{l+1}}(\mathbf{R}) = \phi^{\mathsf{l}}(\mathbf{R}) + \mu \cdot \mathcal{F}^{-1}_\mathbf{k\rightarrow R} \left[\frac{\mathbf{q}\cdot\mathcal{F}_\mathbf{R\rightarrow k}\{\mathbf{I}(\mathbf{R})\}}{ \lambda_1 + k^2 + \lambda_2k^4}\right]\,,
\end{equation*}
with $\mu$ a gradient step size, set here to \num{0.5}, and $\lambda_1$, $\lambda_2$ regularization parameters for lowpass- and highpass-filtering. Fig. 3 shows example reconstructions of a TaTe-filled multi-wall carbon nanotube recorded with the 4D Camera. The datasets with \num{512 x 512} probe positions were recorded within \SI{3}{\second} at a beam energy of \SI{80}{\kilo\electronvolt} with a real-space sampling of \SI{31}{\pico\meter}, and a beam current of \SI{42}{\pico\ampere}.
\begin{figure*}[h]
    \centering
        \includegraphics[width=10cm]{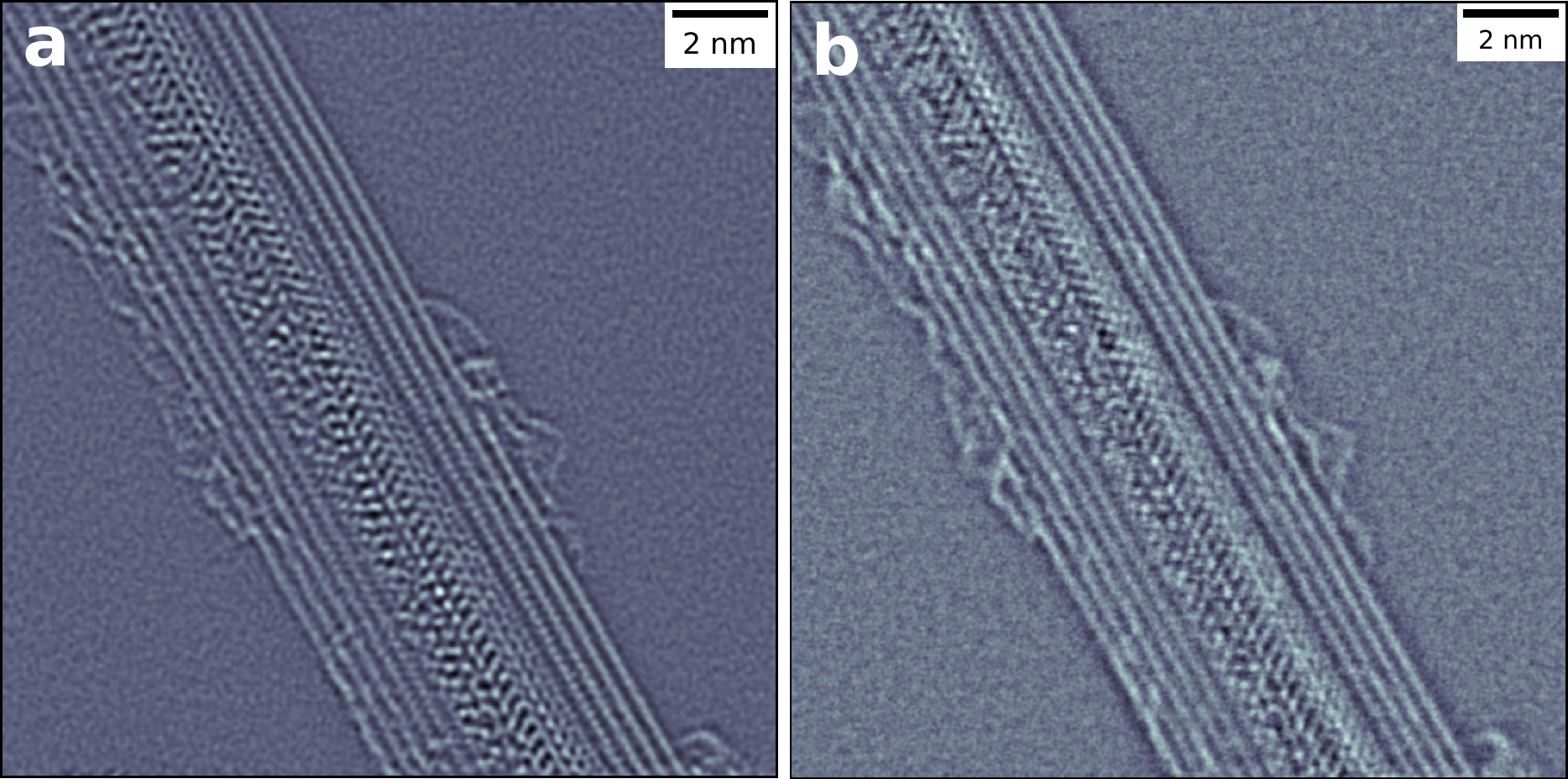}
    \caption{a) Real-time ptychographic reconstruction of a multi-wall carbon-nanotube filled with a complex TaTe structure from 4D-STEM data in EER format. b) DPC reconstruction from 4D-STEM data in EER format.}
	\label{fig:ssb_recon}
\end{figure*}
\section{Conclusion}
We have described  and demonstrated a real-time implementation of SSB ptychography and DPC based on a linear index EER for fast phase-contrast imaging from large 4D-STEM datasets. While many other algorithms for phase reconstruction of 4D-STEM data are available, we have focused here on the ones that can be implemented to reconstruct a phase image in sub-second time, either by a single pass through the 4D dataset, or by fast reduction to 2D, with memory requirements that allow execution on a single GPU. This alleviates the operational complexities of performing low-dose 4D-STEM phase-contrast experiments and enables immediate assessment of the data quality and experimental parameters like residual aberrations.
Future improvements to the detector will allow sub-pixel determination of electron event-coordinates, leading to improved data quality, and electron counting and on-the-fly data reduction on field-programmable gate arrays. Future additions to the real-time reconstruction code could include real-time denoising, e.g. by real-time GPU implementations of the BM3D algorithm, or by real-time neural network alternatives \cite{Burger_Schuler_Harmeling_2012,Gharbi_Chaurasia_Paris_Durand_2016}.
We expect electron event representations to gain in significance in 4D-STEM because of the computational and storage advantages described here. All source code is available as jupyter notebooks at \href{https://github.com/PhilippPelz/realtime_ptychography}{this github repository} and in the py4DSTEM package \cite{Savitzky_Hughes_Zeltmann_Brown_Zhao_Pelz_Barnard_Donohue_DaCosta_Pekin_2020} upon publication.
\section*{Acknowledgment}
We would like to thank Peter Denes, Andrew M. Minor, James Ciston, and John Joseph who contributed to the development of the 4D Camera. We thank Scott Stonemeyer for providing the multi-wall carbon nanotube samples.
\ifCLASSOPTIONcaptionsoff
  \newpage
\fi
\bibliographystyle{unsrt}
\bibliography{bare_jrnl}

\begin{IEEEbiography}[{\includegraphics[width=1in,height=1.25in,clip,keepaspectratio]{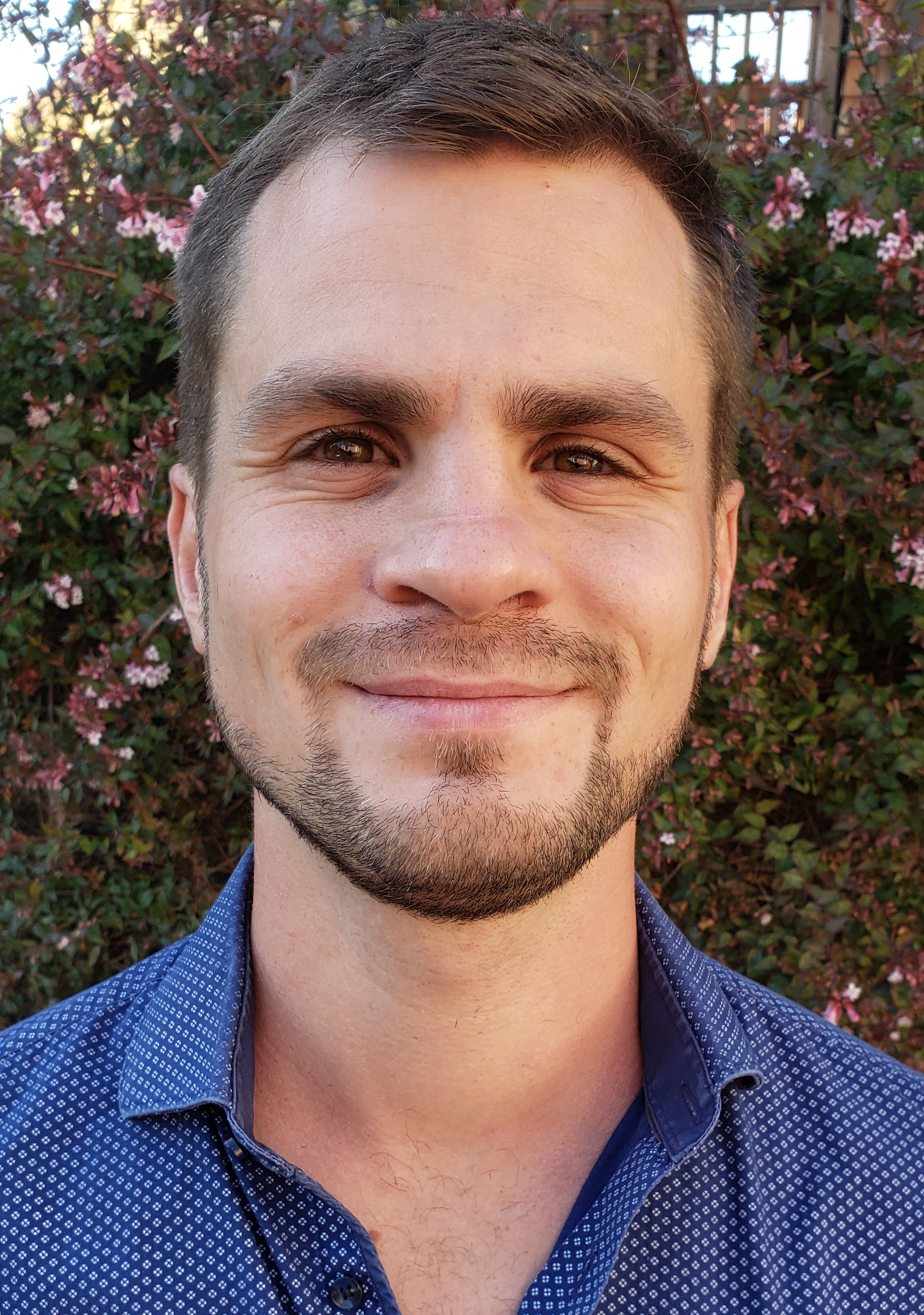}}]{Philipp M. Pelz}
obtained a Ph.D. in Physics from the Max Planck Institute for the Structure and Dynamics of Matter in Hamburg, Germany. He is a postdoctoral researcher in the Materials Science \& Engineering Department at UC Berkeley. His research interests include computational imaging, phase-contrast electron tomography, and design of large-scale optimization algorithms.
\end{IEEEbiography}

\begin{IEEEbiography}[{\includegraphics[width=1in,height=1.25in,clip,keepaspectratio]{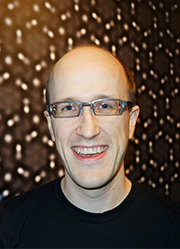}}]{Colin Ophus}
Colin Ophus received his PhD in Materials Engineering from the University of Alberta in Canada. He is currently a staff scientist at the National Center for Electron Microscopy, part of the Molecular Foundry at Lawrence Berkeley National Laboratory. He primarily works on developing methods, algorithms, and codes for simulation, analysis, and instrument design for high resolution and scanning transmission electron microscopy.
\end{IEEEbiography}

\begin{IEEEbiography}[{\includegraphics[width=1in,height=1.25in,clip,keepaspectratio]{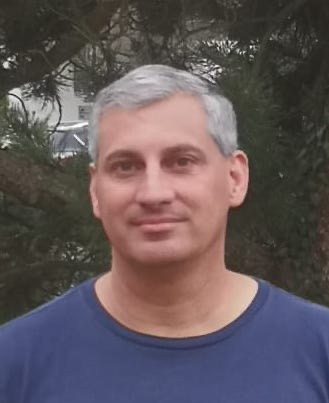}}]{Ian Johnson} obtained his Ph.D. in Physics from the University of California, Davis. He has contributed to a variety of high frame rate detector systems for X-ray Science and Electron Microscopy as a staff scientist at Paul Scherrer Institute (2008-2014) then at Berkeley Lab (2014-2021). His interests include developing real--time instrumentation, data acquisition systems and high--throughput data processing devices for science and technology.
\end{IEEEbiography}

\begin{IEEEbiography}[{\includegraphics[width=1in,height=1.25in,clip,keepaspectratio]{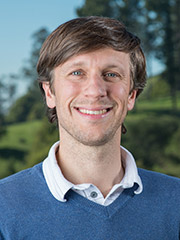}}]{Peter Ercius}
obtained his Ph.D. from Cornell University in 2009. He was a postdoctoral researcher at the National Center for Electron Microscopy from 2009 - 2011 and then joined the staff as a staff scientist. His research interests include high-resolution imaging, electron tomography, \emph{in situ} liquid cell TEM.
\end{IEEEbiography}

\begin{IEEEbiography}[{\includegraphics[width=1in,height=1.25in,clip,keepaspectratio]{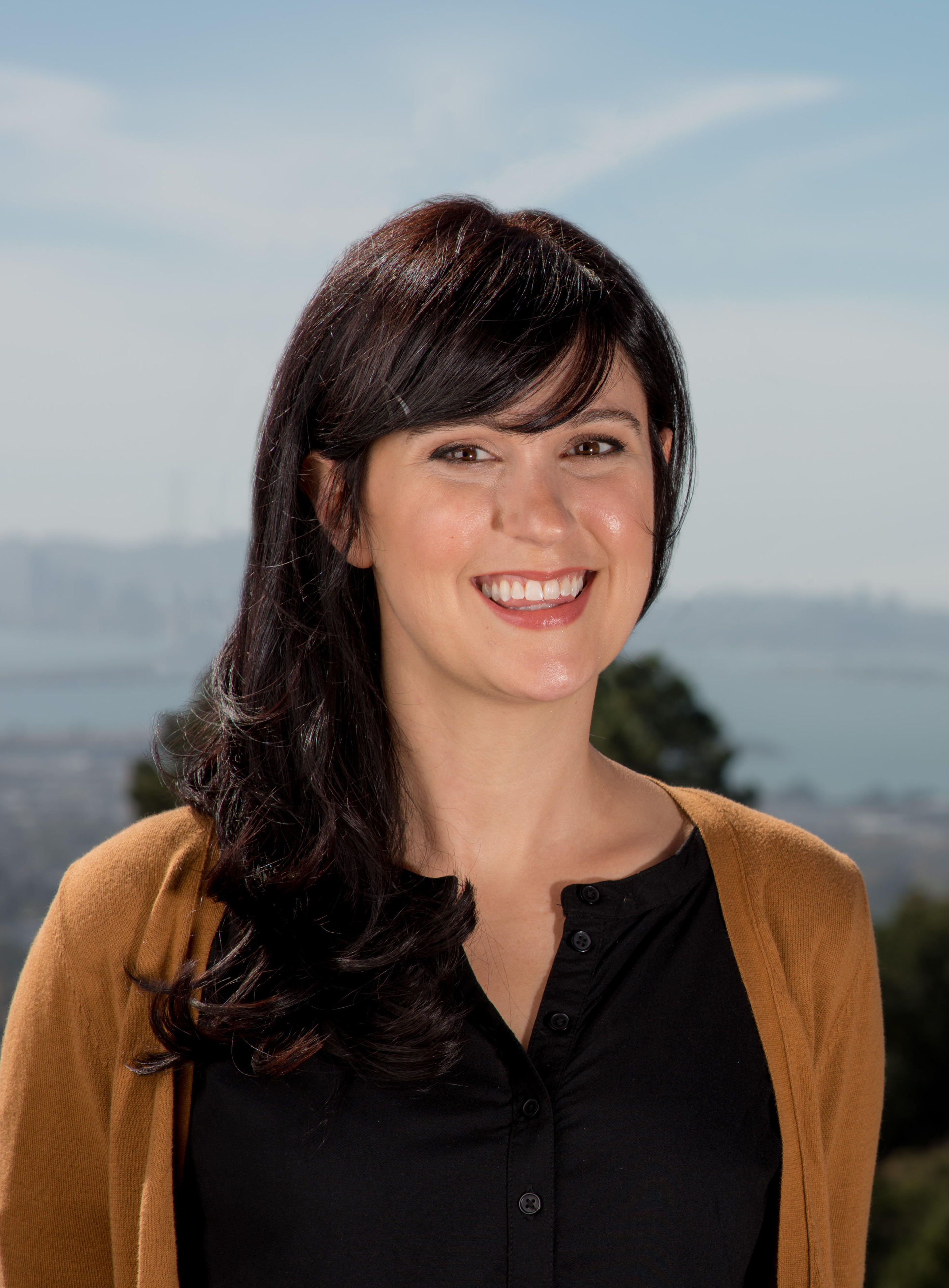}}]{Mary Scott}
obtained her Ph.D. in Physics from the University of California, Los Angeles in 2015. She was a postdoctoral researcher at the University of California, Berkeley until 2017. She is currently an assistant professor in the Materials Science and Engineering department at the University of California, Berkeley as well as a Faculty Staff Scientist at the National Center for Electron Microscopy, part of the Molecular Foundry at Lawrence Berkeley National Lab. Her research interests include electron tomography, high throughput electron microscopy, and electron scanning nanodiffraction.
\end{IEEEbiography}




\end{document}